# Finite Size Effect on Nanomechanical Mass Detection: Role of Surface Elasticity

Mai Duc Dai[1], Chang-Wan Kim[1,*], and Kilho Eom[2,*]

[1]School of Mechanical Engineering, Konkuk University, Seoul 143-701, Republic of Korea
[2]Department of Mechanical Engineering, Korea University, Seoul 136-701, Republic of Korea

*Correspondence should be addressed to C.-W.K. (E-mail: goodant@konkuk.ac.kr) and K.E. (E-mail: kilhoeom@korea.ac.kr).

**Abstract**
Nanomechanical resonators have recently been highlighted because of their remarkable ability to perform the sensing and detection. Since the nanomechanical resonators are characterized by large surface-to-volume ratio, it is implied that the surface effect plays a substantial role on not only the resonance but also the sensing performance of nanomechanical resonators. In this work, we have studied the role of surface effect on the detection sensitivity of a nanoresonator that undergoes either harmonic vibration or nonlinear oscillation based on the continuum elastic model such as beam model. It is shown that surface effect makes an impact on both harmonic resonance and nonlinear oscillations, and that the sensing performance is dependent on the surface effect. Moreover, we have also investigated the surface effect on the mechanical tuning of resonance and sensing performance. It is interestingly found that the mechanical tuning of resonance is independent of surface effect, and that the mechanical tuning of sensing performance is determined by surface effect. Our study sheds light on the importance of surface effect on the sensing performance of nanoresonators.

## I. Introduction

Nanomechanical resonators have recently been received significant attentions for their ability to perform specific, notable functions such as sensing and/or actuations [1-4]. Specifically, nanomechanical resonators have enabled not only the fundamental understanding of vibration theory [5-7] but also the detection and/or measurement of physical quantities such as quantum state [8-10], force [11-13], mass [14-17], surface stress [18-21], surface elastic stiffness [22, 23], biomolecular interactions [24-28], fluid-solid interactions [24, 29-31], and so forth. The remarkable performance of nanoresonators relies on their high-frequency dynamics even up to Mega- or Giga-Hertz [32-35], which could be achieved by scaling down of a resonator. This implies that micro- or nanoscale resonators allow the realization of lab-on-a-chip but also the development of high-frequency devices such as radio communication devices [36]. In addition, scaling down of a resonator increases the detection sensitivity, which results in nanoresonators to be a sensitive mass sensor [14-16]. In particular, based on continuum elasticity [37, 38], the sensing performance is given by $S = \Delta f/\Delta m = (f/2m)$[37, 38], where $\Delta f$ is the resonant frequency shift for a resonator due to the adsorption of mass $\Delta m$, and $f$ and $m$ represent the resonant frequency and the mass of a bare resonator (i.e. without mass adsorption), respectively. This clearly elucidates that the scaling down of a resonator increases the resonant frequency, and consequently amplifies the resonant frequency shift due to mass adsorption. This principle has led a couple of research groups to the development of NEMS resonator for label-free detection of gas atoms [14, 39] and/or gold atoms [15, 16] even in atomic resolution, and a single protein molecule [40]. This suggests that nanomechanical resonators envisage the novel development of lab-on-a-chip mass spectrometers [41].

For effective design of nanomechanical resonators for their specific functions such as detection, it is essential to understand and predict the dynamic behavior of a nanoresonator and its sensing performance. Most previous works [7, 33, 39, 42-44] have analyzed the experimentally measured resonance behavior of nanomechanical resonators using continuum mechanics model (i.e. elastic beam model), which assumes that the resonance of a nanoresonator is attributed to the flexural motion. Despite widely utilized for analysis of resonance, conventional continuum mechanics models fail to characterize the dynamic behavior of a nanoresonator, especially a resonator whose transverse dimension (i.e. diameter) is smaller than sub-50 nm, at which surface effect plays a significant role in the mechanical deformation of a nanostructure [45]. Specifically, when a device is scaled down, the surface energy $U_S$ defined as energetic cost to create a surface due to scaling down increases so as to generate the surface stress $\tau$, which is defined as $\tau = \partial U_s/\partial \varepsilon = \tau_0 + S\varepsilon + O(\varepsilon^2)$ [46-48], where $\varepsilon$ is a mechanical strain of a structure, and $\tau_0$ and $S$ indicate the constant surface stress and the strain-dependent surface stress (i.e. surface elastic stiffness), respectively.

The surface stress effect on the mechanics of nanostructures has recently been thoroughly studied. For example, Miller and Shenoy [49] argued that the bending/axial deformation of a nanoscale beam is well depicted by a continuum model that accounts for the surface stress effect. Cuenot et al. [50] experimentally showed that the bending deformation of a one-dimensional nanostructure such as nanowires is significantly affected by the surface stress. Lilley and a coworker [51] have taken into account the surface elasticity-based continuum mechanics model that allows the fundamental insights into the role of surface elasticity on the mechanical properties (i.e. bending deformation) of nanowires. Wang and Li [52] have recently reported that the surface effect significantly determines the elastic properties of nanowires using the density functional theory (DFT) simulations. Yun and Park [53] have provided that, by using multiscale simulations based on Surface Cauchy-Born model [54-56], the bending behavior of fcc metal nanowires is governed by surface stress.

Recently, the surface stress effect on the resonance of NEMS devices has been studied. For instance, Lu et al. [22] have considered a surface elasticity-based beam model in order to understand the resonant cantilever-based sensing mechanism, especially the size effect on sensing performance of a cantilever due to surface effect. In a similar spirit, Lilley and a coworker [57] have studied the surface stress effect on the resonance behavior of a nanoscale beam using surface elasticity-based beam model. Recent experimental and theoretical studies [18, 21, 58] reported that when biomolecules are adsorbed onto micro- or nanocantilever, the resonance behavior is significantly affected by surface stress due to adsorbed biomolecules rather than added molecular mass, albeit the elastic beam model employed in these studies [18, 21, 58, 59] contradicts the Newton's third law as discussed in classical work by Gurtin et al. [60] and recent studies by Lu et al. [22] and Sader et al. [20]. Furthermore, Park and coworkers [61, 62] have extensively studied the resonance behavior of various nanowires by using multiscale simulations. They have found that the surface stress effect is a key parameter that determines the resonance of nanostrcutures.

It has to be noted that previous studies [20, 22, 57, 63, 64] have considered the surface elasticity-based beam model, which is restricted to the harmonic oscillation. However, it is necessary to gain a fundamental insight into nonlinear vibration of nanoresonators, because nanoresonators can easily undergo the nonlinear oscillation [7, 43, 44]. More importantly, nonlinear vibration of a nanoresonator has recently been highlighted because of unique dynamic characteristics such as frequency tuning [43], chaotic dynamics [7], synchronization [65], and so forth. Moreover, Buks et al. [66] have theoretically studied the detection limit for mass sensing using a nanoresonator that can undergo the nonlinear oscillation. Recently, Eom and coworkers [67] have computationally studied the detection sensitivity of a nanoresonator that is operated in either harmonic oscillation or nonlinear vibration. It is remarkably shown that nonlinear vibration due to doubly clamped boundary condition increases not only the resonant frequency but also the detection sensitivity. In addition, it is suggested that the detection principle for nonlinear vibration-based sensing is very unique. In particular, the mass adsorption onto a nanoresonator undergoing the nonlinear vibrations increases the resonant frequency, which is contrary to the detection principle for the harmonic resonance-based detection. Furthermore, in ref. [67], it is found that a mechanical force can amplifies the detection sensitivity for a nanoresonator enduring the harmonic oscillation, whereas a mechanical force significantly reduces the detection sensitivity for a nanoresonator sustaining the nonlinear oscillation. These examples show that nonlinear oscillation can be a unique route to improve the detection sensitivity for a nanomechanical resonator, albeit the nonlinear oscillation-based mass sensing has been rarely taken into account.

In this work, we are aimed at studying the role of surface stress effect on the nonlinear vibration-based mass detection. In particular, we have studied the resonance behavior of silicon nanowire resonators that can experience either harmonic oscillation or nonlinear vibration using a surface elasticity-based beam model. We have found that the surface effect makes an impact on not only the (harmonic or nonlinear) vibrations but also the sensing performances based on both harmonic and nonlinear resonances. Further, we have shown that surface effect does also play a substantial role on the mechanical tuning of resonance as well as the mechanical tuning of sensing performance. The remainder of this article consists as follows: Section II describes the surface elasticity-based elastic beam model for silicon nanowires. Section III provides the numerical simulations of the vibration of silicon nanowires, the sensing performance, and the mechanical tuning of resonance or sensing performance. Section IV encloses our article with remarks.

## II. Theory and Model

We have modeled a nanoresonator as a one-dimensional elastic beam, because nanoresonators exhibit the dimension such that the transverse dimensions (i.e. thickness and width) are much smaller than the longitudinal dimension (i.e. length). Here, we consider the silicon nanowires for nanoresonators, since recent studies [42, 68] have reported that doubly-clamped silicon nanowires can be actuated by magnetomotive technique [42] and/or piezoresitive method [68].

It is noted that since nanoresonators are characterized by large surface-to-volume ratio, the vibrational behavior of nanoresonators is governed by not only the strain (bending) energy $U_b$ but also the surface energy $U_s$ defined as the energetic cost to create a surface due to scaling down. This indicates that the surface stress effect is a key factor that determines the vibration of nanoresonator. It is implied that in order to model a nanoresonator, the surface stress effect has to be incorporated with a classical beam model, which is demonstrated as below.

In order to model a nanoresonator, we consider the elastic beam that inherently exhibits the surface stress, because the elastic properties of nanowires with their transverse dimension of <50 nm are significantly affected by surface stress as described in ref. [45]. Let us introduce the surface stress $\tau$ represented in the form of $\tau = \tau_0 + S\varepsilon$ [46-48], where $\tau_0$ is the constant surface stress, $S$ is the strain-dependent surface stress (i.e. surface elastic stiffness), and $\varepsilon$ is the mechanical strain. For silicon nanowires fabricated along the (100) crystallographic direction, the elastic modulus and the surface elastic stiffness are given by $E = 179$ GPa and $S = -18.1$ N/m [49], respectively. Here, the cross-sectional shape of a silicon nanowire is assumed as a hexagonal shape. It should be noted that when surface stress is exerted on an elastic body, the residual stress $\sigma_R$ appears in order to satisfy the equilibrium condition from Newton's third law. Specifically, we have a relation between surface stress and residual stress from force equilibrium:

$$b\int_{-d/2}^{d/2} \tau \cdot \delta(y \pm t/2) dy + \frac{4}{\sqrt{3}} \int_{-d/2}^{d/2} \tau \, dy + \iint_A \sigma_R \, dA = 0 \qquad (1)$$

where $b$ and $d$ are the width and the thickness of a nanoresonator, respectively, $\delta(y)$ is the Dirac delta function, and $y$ is a coordinate along the thickness. When an elastic beam that exerts the surface stress undergoes the bending deformation, the bending moment for such a beam is given by

$$M = b\int_{-d/2}^{d/2} y\tau \cdot \delta(t \pm t/2) dy + \frac{4}{\sqrt{3}} \int_{-d/2}^{d/2} \tau y \, dy + \iint_A y(\sigma_b + \sigma_R) dA \qquad (2.a)$$

Here, $\sigma_b$ is the bending stress given by $\sigma_b = EI\kappa$, where $E$ is the bending modulus of a bulk material, $I$ is the cross-sectional moment of inertia, and $\kappa$ is the bending curvature. From Eqs. (1) and (2.a), the bending moment is represented in the form

$$M(x,t) = \left(\frac{Sbd^2}{2} + \frac{Sd^3}{3\sqrt{3}} + EI\right)\frac{\partial^2 w(x,t)}{\partial x^2} \qquad (2.b)$$

Therefore, the equation of motion for a nanoresonator is given by

$$\left(EI + Sd^3/3\sqrt{3} + Sbd^2/2\right)\frac{\partial^4 w(x,t)}{\partial x^4} - \left[T_0 + \frac{EA}{2L}\int_0^L \left\{\frac{\partial w(x,t)}{\partial x}\right\}^2 dx\right]\frac{\partial^2 w(x,t)}{\partial x^2}$$
$$+ \gamma\frac{\partial w(x,t)}{\partial t} + \xi\frac{\partial^2 w(x,t)}{\partial t^2} = f(x,t) \qquad (3)$$

where $A$ and $L$ represent the cross-section area and the length of a nanoresonator, respectively, $\gamma$ is the damping coefficient, $T_0$ is the mechanical tension applied to a resonator, $\xi$ is the effective mass per unit length for a nanoresonator, and $f(x, t)$ is the actuation force per unit length. Here, the effective mass per unit length for a bare resonator without mass adsorption is given as $\xi = \rho A$, where $\rho$ is the mass density of a resonator. In addition, the actuation force per unit length is assumed to be in the form of $f(x, t) = F(x)\cdot\cos(\Omega t)$, where $\Omega$ is the driving frequency. It should be noted that the doubly-clamped boundary conditions induce the geometric nonlinear deformation dictated by a term $(EA/2L)\int_0^L (\partial w/\partial x)^2 dx$ [69, 70]. For solving the equation of motion depicted by Eq. (3), we have employed the Galerkin's method [70] that assumes the bending deflection $w(x, t)$ in the form of $w(x, t) = z(t)\cdot\psi(x)$, where $z(t)$ is the time-dependent amplitude and $\psi(x)$ is the deflection eigenmode. Here, we assume the deflection eigenmode given as $\psi(x) = (2/3)^{1/2}[1 - \cos(2\pi x/L)]$ that is the deflection eigenmode for a vibrating beam without considering any geometric nonlinear effect and mass adsorption. Here, it should be noted that the deflection eigenmode $\psi(x)$ satisfies the essential boundary conditions, i.e. $\psi(0) = \psi(L) = \psi''(0) = \psi''(L) = 0$. By multiplication of deflection eigenmode $\psi(x)$ with Eq. (3) followed by integration by parts, the equation of motion becomes the Duffing equation [71-73] such as

$$(\alpha + \beta T_0)z(t) + \lambda[z(t)]^3 + \eta\dot{z}(t) + \mu\ddot{z}(t) = p\cos\Omega t \qquad (4)$$

where parameters are given as

$$\alpha = \left(EI + \frac{Sbd^2}{2} + \frac{Sd^3}{3\sqrt{3}}\right)\int_0^L (d^2\psi/dx^2)^2 dx = \frac{16\pi^4}{3L^3}\left(EI + \frac{Sbd^2}{2} + \frac{Sd^3}{3\sqrt{3}}\right) \qquad (5.a)$$

$$\beta = \int_0^L (d\psi/dx)^2 dx = \frac{4\pi^2}{3L} \qquad (5.b)$$

$$\lambda = \frac{EA}{2L}\left[\int_0^L (d\psi/dx)^2 dx\right]^2 = \frac{8\pi^4 EA}{9L^3} \qquad (5.c)$$

$$\eta = \gamma\int_0^L \psi^2 dx = \gamma L \qquad (5.d)$$

$$\mu = \int_0^L \xi\psi^2 dx = \rho AL + R \qquad (5.e)$$

$$p = \int_0^L F(x)\cdot\psi(x) dx \qquad (5.f)$$

Here, $R$ is the effective mass for added atoms, and $p$ can be regarded as a total amount of force that is used to actuate the nanoresonator. For mass adsorption that occurs uniformly over the entire length $L$, the effective mass per unit length is given as $\xi = \rho A + \Delta M/L$, where $\Delta M$ is the total amount of added mass. For such a case, a term $R$ is given by $R = \Delta M$. In case of mass adsorption that happens locally at the location of $x = a$ (where $0 < a < L$), the effective mass per unit length becomes $\xi = \rho A + \Delta M \cdot \delta(x - a)$, where $\delta(x)$ is the Dirac delta function. As a consequence, for the localized mass adsorption, a term $R$ is in the form of $R = \Delta M[1 - \cos(2\pi a/L)]$. The resonance behaviors of nanoresonators as well as their response to mass adsorption are numerically studied by solving the Duffing equation given by Eq. (4).

## III. Results and Discussion

### A. Resonance Behaviors of Nanoresonators: Role of Surface Effect

We have taken into account the resonance behavior of silicon nanowires that can undergo either harmonic oscillations or nonlinear vibrations based on numerical simulations of Duffing equation given by Eq. (4). First, we consider the harmonic oscillations of silicon nanowires in order to compute the effective elastic modulus of silicon nanowires, because the harmonic resonance-based method [74, 75] has been widely utilized for measuring the elastic modulus of nanowires. In addition, the elastic properties of silicon nanowires have recently been extensively studied with respect to their sizes such as cross-sectional dimension [45, 76-78]. For harmonic oscillation, with neglecting the geometric nonlinear effect, the resonant frequency of a doubly-clamped nanowire is given by [70]

$$\omega_i = \left(\frac{\kappa_i}{L}\right)^2 \sqrt{\frac{E_{eff} I}{\rho A}} \qquad (6)$$

where $\omega_i$ is the resonant frequency for the $i$-th mode, $\kappa_i$ is the $i$-th eigenvalue for doubly-clamped boundary condition, i.e. $\kappa_1 = 4.73$ (for fundamental resonance), $\kappa_2 = 7.85$ (for the second harmonic resonance), etc., and $E_{eff}$ is the effective elastic modulus for a silicon nanowire. In order to consider the harmonic resonance, the Duffing equation is taken into account with small amount of actuating force $p$ such as $p = 0.1$ aN (where 1 aN = $10^{-18}$ N). The inset of Figure 1 shows the resonance curve for a silicon nanowire whose diameter and length are given as $d = 20$ nm and $L = 2.2$ μm. It is clearly shown that, with small amount of $p$, i.e. $p = 0.1$ aN, the resonance behavior is well fitted to harmonic oscillation. From the measured resonant frequency, we can calculate the effective elastic modulus from Eq. (6). Figure 1 shows the effective elastic modulus of various silicon nanowires whose diameters are in the range of 10 nm to 70 nm. It is interestingly found that as the diameter of a silicon nanowire decreases, the effective elastic modulus is significantly reduced. This is attributed to the negative surface elastic stiffness for silicon fabricated along (100) crystallographic direction. Our numerical results on the dependence of effective elastic modulus of silicon nanowire on its diameter are consistent with recent experimental and computational observations [45, 76, 77, 79] that the effective modulus of a silicon nanowire decreases when its diameter is reduced. In particular, as shown in Fig. 1, our theoretical prediction on the size dependence of elastic modulus for silicon nanowire is quantitatively comparable to first-principle calculations [79] on elastic modulus. Moreover, the size dependence of elastic modulus predicted from our theoretical model is also comparable to that obtained from experiment [76], albeit our theoretical model overestimates the elastic modulus of nanowires with their diameter less than 20 nm (Fig. 1). The discrepancy between theoretical prediction and experimental observation for elastic modulus of silicon nanowire with its diameter less than 20 nm may be attributed to the surface effect-driven phase transition, which may significantly affect the atomic structure of nanowire and consequently the elastic properties of nanowires [80]. It is implied that our theoretical model based on the continuum mechanics is appropriate for predicting the mechanical properties of nanowires, as long as the phase transition that makes a critical impact on the atomic structure of nanowire does not occur.

Because our previous study [67] shows that nonlinear oscillation can be a useful route to improving the detection sensitivity for a nanoresonator, it is essential to study the nonlinear oscillations of silicon nanowires for further applications in mass sensing. To our best knowledge, even though recent studies by Lu et al. [22] and Lilley et al. [57] provide the fundamental insights into the role of surface effect (i.e. surface stress) on the resonance behavior of a one-dimensional nanostructure, their studies [22, 57] are restricted to the harmonic oscillations of a 1-D nanostructure (e.g. nanocantilever, and nanowire). In order to induce the nonlinear oscillation, we consider the actuating force $p > 1$ pN (where 1 pN = $10^{-12}$ N). In particular, the resonance behaviors of a silicon nanowire actuated by a force $p = 2.5$ pN or 0.1 fN (where 1 fN = $10^{-15}$ N) are shown in the inset of Figure 2. It is shown that the actuation force of $p = 0.1$ fN induces the harmonic oscillation of a silicon nanowire, while the nonlinear vibration was observed when the actuation force is given as $p = 2.5$ fN. As shown in the inset of Figure 2, the nonlinear oscillation increases the resonant frequency, which is attributed to the geometric nonlinear effect due to doubly-clamped boundary condition. In order to understand the role of surface effect as well as the finite size effect on the nonlinear vibrations, we take into account the resonant frequencies of silicon nanowires actuated by an excitation force $p$ in the range of 0.5 nN to 2.5 nN (Figure 2). For envisaging the role of surface effect, we have considered two models – (i) a continuum model with excluding the surface effect, and (ii) continuum model that accounts for the surface effect. The dashed lines in Figure 2 show the resonant frequencies calculated from a model that excludes the surface effect, while solid lines indicate the resonant frequencies estimated from a continuum model that considers the surface effect. As actuation force $p$ increases, the normalized resonant frequencies are critically dependent on the surface effect. Here, the resonant frequencies are normalized by the harmonic resonances. Furthermore, as surface-to-volume ratio increases, the surface effect significantly dominates the resonance behavior of nanoresonators (Figure 2). In summary, for a nanoresonator that exhibits the large surface-to-volume ratio, the nonlinear vibration behavior is substantially determined by the surface effect (i.e. surface elastic stiffness).

## *B. Surface Effect on Resonator-Based Mass Sensing*

Because nanoresonators have recently been highlighted as a highly sensitive mass detector [14-16], it is essential to study the dynamic response of nanoresonators to mass adsorption. As delineated above, the surface effect plays a substantial role in the resonant frequencies of nanoresonators, which implies that the detection sensitivity for a nanoresonator may be highly correlated with the surface effect. This section is aimed at gaining insight into the role of the surface effect on the sensing performance of nanoresonators that can undergo either harmonic or nonlinear oscillations.

First, we consider the case where the mass adsorption occurs on a nanoresonator that experiences the harmonic resonance. The inset of Figure 3 depicts the resonance curves for a bare silicon nanowire (with diameter $d = 10$ nm) and a silicon nanowire with mass adsorption (i.e. mass of 80 ag). It is shown that when a silicon nanowire is actuated by an actuation force $p = 0.1$ fN, the resonant frequency for a silicon nanowire is decreased due to added mass. We have investigated the resonant frequency shifts due to mass adsorption as a function of surface-to-volume ratio as well as the amount of adsorbed mass. It is found that the resonant frequency shift due to mass adsorption is dependent on the surface-to-volume ratio such that the larger surface-to-volume ratio of a nanoresonator, the better detection sensitivity does it possess. Furthermore, as shown in Figure 3, the resonant frequency shift due to mass adsorption is linearly proportional to the amount of adsorbed mass, which is consistent with the detection principle for harmonic oscillation-based detection. For an insight into the surface effect on the sensing performance, we take into account the difference between resonant frequencies for a nanoresonator with mass adsorption, which were calculated from two models – (i) a continuum model that excludes the surface effect, and (ii) a continuum model that accounts for the surface effect. Specifically, the frequency difference is defined as $\Delta\Omega = \omega - \Omega$, where $\omega$ is the resonant frequency for a resonator with mass adsorption is estimated from a continuum model that considers the surface effect, while $\Omega$ is the resonant frequency is calculated from a model that excludes the surface effect. Here, the resonance behavior is only restricted to harmonic oscillation. It is shown that when very small amount of mass (e.g. mass of 10 ag) is adsorbed onto a resonator, the frequency difference is miniscule and independent of surface-to-volume ratio (Figure 4). This indicates that when the adsorption of small amount of mass occurs, the sensing performance of a nanoresonator undergoing harmonic resonance is uncorrelated with the surface effect. On the other hand, as shown in Figure 4, it is found that when the adsorption of sufficient amount of mass (i.e. mass of ≥ 20 ag), the frequency difference between two models is critically dependent on both the surface-to-volume ratio and the amount of adsorbed mass. In particular, the frequency difference is almost linearly proportional to the surface-to-volume ratio. This indicates that in case of the adsorption of mass of ≥ 20 ag onto a silicon nanowire that exhibits the surface-to-volume ratio of >0.2 $nm^{-1}$, the surface effect takes a significant role in the sensing performance of a nanoresonator.

As described in Section III.A, the nonlinear oscillation behavior of a nanoresonator is significantly affected by the surface effect. Moreover, our previous study [67] reports that the detection sensitivity for a resonator-based mass sensing could be remarkably enhanced by using nonlinear oscillation. These led us to study the role of surface effect on the detection sensitivity for nonlinear resonance-based detection. As elucidated in our previous report [67], the mass adsorption increases the resonant frequency for a nanoresonator operated in nonlinear oscillation (see also an inset of Figure 5), which is quite different from the detection principle for harmonic oscillation-based sensing. We have investigated the size effect on the resonant frequency shift for a resonator undergoing the nonlinear vibration due to mass adsorption. As shown in Figure 5, the resonant frequency shift due to mass adsorption depends on the surface-to-volume ratio. In order to gain the fundamental insights into the role of surface elasticity on the nonlinear oscillation-based mass sensing, we have taken into account the difference between frequencies computed from two models – (i) a model that excludes the surface effect, and (ii) a continuum model that accounts for the surface effect. Specifically, the frequency difference is defined as $\Delta\Omega = \omega - \Omega$, where $\omega$ is the resonant frequency for a resonator with mass adsorption is estimated from a continuum model that considers the surface effect, while $\Omega$ is the resonant frequency is calculated from a model that excludes the surface effect. It is found that regardless of the amount of adsorbed mass, the frequency difference is only governed by the surface-to-volume ratio (Figure 6). In other words, unlike the case of mass sensing based on harmonic oscillation, the surface effect makes a substantial impact on the nonlinear resonance-based mass detection regardless of the amount of added mass onto a resonator. Here, it should be noted that the frequency difference $\Delta\Omega$ for nonlinear oscillation is two orders of magnitude larger than that for harmonic oscillation. This is attributed to the fact that large actuation force $p$, i.e. $p = 2.5$ nN, (that leads to the nonlinear oscillation) significantly increases the resonant frequency by two orders of magnitude. It is also noted that the frequency difference $\Delta\Omega$ for harmonic oscillation is obtained based on the actuation force of $p = 0.1$ fN (ensuring the harmonic resonance), which is six orders of magnitude smaller than the actuation force ($p = 2.5$ nN) that is used to drive the nonlinear oscillation.

*C. Mechanical Tuning of Resonance and Sensing Performance: Role of Mechanical Force*

It is fruitful to study the effect of a mechanical force applied to a resonator on the sensing performance of a resonator, since it has recently been found that mechanical force is an efficient route to enhance both the resonant frequency and the quality (Q) factors of a nanoresonator [67, 81-84].

First, let us consider the tuning of resonant frequencies using a mechanical force. In particular, the resonant frequency of a silicon nanowire that undergoes the harmonic oscillation can be tuned by application of a mechanical force $T_0 = 0.6$ nN such that a mechanical force increases the resonant frequency (see the inset of Figure 7). This can be easily elucidated from a continuum mechanics model. For small deformation with neglecting the damping effect, i.e. $\lambda \approx 0$ and $\eta \approx 0$, the resonant frequency under application of a mechanical tension $T_0$ is given by $\omega = \omega_0[1 + T_0L^2/(\alpha EI)]^{1/2}$ [67, 81, 86], where $\omega_0$ is the resonant frequency for a resonator without any application of mechanical force, and $\alpha$ is a boundary condition-dependent constant, i.e. $\alpha = 4\pi^2/0.97$ for doubly-clamped boundary condition [86]. As a consequence, the resonant frequency shift $\Delta\omega$ due to applied mechanical force $T_0$ is represented in the form of $\Delta\omega/\omega_0 = \varepsilon/2 + O(\varepsilon^2)$, where $\varepsilon$ is a normalized mechanical tension defined as $\varepsilon = T_0L^2/(\alpha EI)$. This is consistent with our numerical result (i.e. Figure 7) that shows the normalized frequency shift due to applied mechanical tension $T_0$ for a silicon nanowire resonator undergoing the harmonic oscillation. Specifically, the frequency shift due to mechanical tension is almost linearly proportional to the amount of applied mechanical force. Moreover, it is found that the normalized frequency shift due to mechanical tension depends on the surface-to-volume ratio (Figure 7). In order to investigate of the correlation between surface effect and mechanical tuning of resonant frequency, we have taken into account a model that excludes the surface effect in order to compute the resonant frequency shift due to mechanical tension (shown as dashed lines in Figure 7). It is shown that for a nanoresonator that exhibits the surface-to-volume ratio of $\leq 0.4$ nm$^{-1}$, the resonant frequency shift due to mechanical tension is almost independent of the surface effect.

Furthermore, we have studied the mechanical tuning of nonlinear resonances using a mechanical force. Unlike the case of harmonic oscillation, the application of a mechanical force to a nanoresonator operated in nonlinear vibrations reduces the resonant frequency (see the inset of Figure 8). It is shown that for nanoresonators that possess the surface-to-volume ratio of $\leq 0.2$ nm$^{-1}$, a decrease in the resonant frequencies due to mechanical tension is linearly proportional to the applied mechanical force $T_0$ (Figure 8). On the other hand, for a resonator that has the surface-to-volume ratio of $>0.2$ nm$^{-1}$, the frequency shift due to mechanical tension is no longer linearly proportional to the mechanical force $T_0$. In addition, in order to gain insights into the surface effect on the mechanical tuning of nonlinear resonances, we have considered a continuum model that excludes the surface effect. The frequency shift due to mechanical tension, which is computed from such a model (i.e. neglecting the surface effect), is almost close to that estimated from the model that accounts the surface effect (Figure 8). This indicates that the surface elastic stiffness does not significantly affect the resonant frequency shift due to a mechanical tension. In other words, the mechanical tuning of nonlinear resonances is insensitive to the surface effect.

Now, let us study the mechanical tension-based tuning of sensing performance, because a mechanical tension is a useful avenue to tune the sensing performance of nanoresonators [67, 83]. Here, we have measured the frequency shift for a resonator, which bears a mechanical tension, due to mass adsorption based on our theoretical model depicted in Eq. (4). First, we consider the case of harmonic oscillation for mass sensing using a resonator that exerts a mechanical tension. Figure 9 shows the resonant frequency shift due to mass adsorption (with amount of 40 ag) for a resonator to which a mechanical tension is applied. It is interestingly found that frequency shift is strongly dependent on not only the amount of mechanical tension but also the surface-to-volume ratio. In particular, for a surface-to-volume ratio of $<0.3$ nm$^{-1}$, a mechanical tension does not make any impact on the resonant frequency shift due to added mass. On the other hand, for the surface-to-volume ratio of $>0.5$ nm$^{-1}$, the resonant frequency shift due to adsorbed mass is sensitively dependent on a mechanical tension. This indicates that the finite size effect is a key factor that determines the mechanical force-based tuning of the detection sensitivity. This can be elucidated from a continuum model that takes into account harmonic resonance with neglecting the geometric nonlinearity and damping effect (i.e. $\lambda \approx 0$ and $\eta \approx 0$ in Eq. (4)). The resonant frequency for a bare resonator, without any added mass, that exerts a mechanical tension $T_0$ is given by $\omega_0 = \varphi(1 + Sbd^2/2EI + 3^{-3/2}Sd^3/EI)^{1/2}[1 + T_0L^2/(\alpha EI)]^{1/2}$, where $\varphi = (\kappa/L)^2(EI/\rho A)^{1/2}$. As a consequence, the frequency shift due to mass adsorption is represented in the form of

$$\Delta\omega = -\frac{\Delta m}{2m}\varphi\left[\left(1+\bar{S}\right)\left(1+\bar{T}\right)\right]^{1/2} \tag{7.a}$$

where

$$\bar{S} = \frac{Sd^3}{EI}\left[2\left(\frac{b}{d}\right)+\frac{1}{2\sqrt{3}}\right] \propto \frac{S}{Ed} \quad \text{and} \quad \bar{T} = \frac{T_0 L^2}{\alpha EI} \tag{7.b}$$

Eq. (7) demonstrates that the frequency shift due to mass adsorption for a resonator bearing a mechanical tension depends on both the surface effect and the mechanical tension. In particular, the frequency shift due to added mass is proportional to the factor of $1+\bar{S}$, which indicates that as the diameter $d$ of a resonator decreases (i.e. the surface-to-volume ratio increases), the frequency shift due to adsorbed mass becomes strongly dependent on an applied mechanical tension. However, as the diameter $d$ increases (i.e. surface-to-volume ratio approaches zero), the frequency shift due to mass adsorption is weakly dependent of applied mechanical tension. This is consistent with our numerical results shown in Figure 9.

Moreover, we have investigated the effect of mechanical tension on the frequency shift due to mass adsorption for a resonator that undergoes the nonlinear vibration. It is clearly shown that mechanical tension reduces the frequency shift due to added mass (Figure 10), which indicates that mechanical tension decreases the detection sensitivity. In particular, when a mechanical tension of <5 nN is applied to a resonator, the frequency shift due to mass adsorption is highly correlated with the surface-to-volume ratio of a resonator. However, when a nanoresonator exerts a mechanical force of >10 nN, the frequency shift due to adsorbed mass is significantly reduced and almost independent of the surface-to-volume ratio. This indicates that the detection sensitivity for nonlinear resonance-based detection can be diminished by application of a mechanical tension to a resonator.

In summary, a mechanical tuning of resonance using a mechanical force is a useful route to enhance the detection sensitivity for harmonic oscillation-based sensing, whereas the mechanical tuning deteriorates the detection sensitivity for nonlinear resonance-based detection.

## IV. Conclusion

In this work, we have studied the finite size effect on the dynamic behavior of nanoresonators as well as their sensing performances. In particular, in order to understand such a finite size effect (or equivalently, the surface effect), we have considered a continuum model that accounts for the surface effect for modeling silicon nanowire resonators. It is shown that our model, which takes the surface effect into account, is able to predict the mechanical properties (e.g. elastic modulus, and resonance) of silicon nanowires. This suggests that the surface effect makes a significant impact on the mechanical properties such as resonances of nanowires. We have also studied the surface effect on the nonlinear vibrations. It was found that as a resonator is scaled down, the surface effect becomes a key factor that determines the nonlinear vibrations. More importantly, our study shows that surface effect takes a leading role in the detection sensitivity for resonator-based sensing based on either harmonic or nonlinear oscillations. Specifically, for nonlinear oscillation-based detection, the sensing performance of a nanoresonator is significantly governed by the surface effect regardless of the amount of adsorbed mass. On the other hand, for harmonic oscillation-based sensing, is the surface effect on the sensing performance becomes sensitively dependent on the amount of added mass. In addition, we have investigated the surface effect on the mechanical force-based tuning of resonances and detection sensitivity. It is found that the surface effect does not play any role on the mechanical tuning of resonance, while the surface effect significantly dominates the mechanical tuning of the detection sensitivity for a resonator undergoing either harmonic or nonlinear resonances. In summary, our study sheds light on the importance of the surface effect on not only the vibration behavior of nanoresonators but also their sensing performance. This implies that the surface effect should be carefully taken into account for designing a nanoresonator for its specific functions such as actuations and/or detections.

**Acknowledgement**

This work was supported by National Research Foundation of Korea (NRF) under Grant No. NRF-2009-0071246 and NRF-2010-0026223 (to K.E.) and NRF-2009-0067895 and NRF-2010-00525 (to C.W.K.).

**References**
[1]  H. G. Craighead, Science 290 (2000) 1532.
[2]  K. L. Ekinci, Small 1 (2005) 786.
[3]  M. L. Roukes, Physics World 14 (2001) 25.
[4]  K. Eom, H. S. Park, D. S. Yoon, and T. Kwon, Phys. Rep. (2011) in press (DOI: 10.1016/j.physrep.2011.03.002)
[5]  S.-B. Shim, M. Imboden, and P. Mohanty, Science 316 (2007) 95.
[6]  M. Sato, B. E. Hubbard, A. J. Sievers, B. Ilic, D. A. Czaplewski, and H. G. Craighead, Phys. Rev. Lett. 90 (2003) 044102.
[7]  R. B. Karabalin, M. C. Cross, and M. L. Roukes, Phys. Rev. B 79 (2009) 165309.
[8]  A. D. O'Connell, M. Hofheinz, M. Ansmann, R. C. Bialczak, M. Lenander, E. Lucero, M. Neeley, D. Sank, H. Wang, M. Weides, J. Wenner, J. M. Martinis, and A. N. Cleland, Nature 464 (2010) 697.
[9]  M. D. LaHaye, O. Buu, B. Camarota, and K. C. Schwab, Science 304 (2004) 74.
[10]  K. C. Schwab and M. L. Roukes, Phys. Today 58 (2005) 36.
[11]  T. D. Stowe, K. Yasumura, T. W. Kenny, D. Botkin, K. Wago, and D. Rugar, Appl. Phys. Lett. 71 (1997) 288.
[12]  H. J. Mamin and D. Rugar, Appl. Phys. Lett. 79 (2001) 3358.
[13]  J. L. Arlett, J. R. Maloney, B. Gudlewski, M. Muluneh, and M. L. Roukes, Nano Lett. 6 (2006) 1000.
[14]  Y. T. Yang, C. Callegari, X. L. Feng, K. L. Ekinci, and M. L. Roukes, Nano Lett. 6 (2006) 583.
[15]  K. Jensen, K. Kim, and A. Zettl, Nat. Nanotechnol. 3 (2008) 533.
[16]  E. Gil-Santos, D. Ramos, J. Martinez, M. Fernandez-Regulez, R. Garcia, A. San Paulo, M. Calleja, and J. Tamayo, Nat. Nanotechnol. 5 (2010) 641.
[17]  M. Zheng, K. Eom, and C. Ke, J. Phys. D: Appl. Phys. 42 (2009) 145408.
[18]  K. S. Hwang, K. Eom, J. H. Lee, D. W. Chun, B. H. Cha, D. S. Yoon, T. S. Kim, and J. H. Park, Appl. Phys. Lett. 89 (2006) 173905.
[19]  S. Kim and K. D. Kihm, Appl. Phys. Lett. 93 (2008) 081911.
[20]  M. J. Lachut and J. E. Sader, Phys. Rev. Lett. 99 (2007) 206102.
[21]  J. Dorignac, A. Kalinowski, S. Erramilli, and P. Mohanty, Phys. Rev. Lett. 96 (2006) 186105.
[22]  P. Lu, H. P. Lee, C. Lu, and S. J. O'Shea, Phys. Rev. B 72 (2005) 085405.
[23]  H. Ibach, Surf. Sci. Rep. 29 (1997) 193.
[24]  T. Y. Kwon, K. Eom, J. H. Park, D. S. Yoon, T. S. Kim, and H. L. Lee, Appl. Phys. Lett. 90 (2007) 223903.
[25]  T. Kwon, K. Eom, J. Park, D. S. Yoon, H. L. Lee, and T. S. Kim, Appl. Phys. Lett. 93 (2008) 173901.

[26] T. Kwon, J. Park, J. Yang, D. S. Yoon, S. Na, C.-W. Kim, J. S. Suh, Y. M. Huh, S. Haam, and K. Eom, PLoS ONE 4 (2009) e6248.
[27] S. Kim, D. Yi, A. Passian, and T. Thundat, Appl. Phys. Lett. 96 (2010) 153703.
[28] K. Eom, T. Y. Kwon, D. S. Yoon, H. L. Lee, and T. S. Kim, Phys. Rev. B. 76 (2007) 113408.
[29] S. Kirstein, M. Mertesdorf, and M. Schonhoff, J. Appl. Phys. 84 (1998) 1782.
[30] A. V. E. Cornelis and E. S. John, J. Appl. Phys. 100 (2006) 114916.
[31] D. M. Karabacak, V. Yakhot, and K. L. Ekinci, Phys. Rev. Lett. 98 (2007) 254505.
[32] H. B. Peng, C. W. Chang, S. Aloni, T. D. Yuzvinsky, and A. Zettl, Phys. Rev. Lett. 97 (2006) 087203.
[33] A. Husain, J. Hone, H. W. C. Postma, X. M. H. Huang, T. Drake, M. Barbic, A. Scherer, and M. L. Roukes, Appl. Phys. Lett. 83 (2003) 1240.
[34] H. W. C. Postma, I. Kozinsky, A. Husain, and M. L. Roukes, Appl. Phys. Lett. 86 (2005) 223105.
[35] V. Sazonova, Y. Yaish, H. Ustunel, D. Roundy, T. A. Arias, and P. L. McEuen, Nature 431 (2004) 284.
[36] K. Jensen, J. Weldon, H. Garcia, and A. Zettl, Nano Lett. 7 (2007) 3508.
[37] T. Braun, V. Barwich, M. K. Ghatkesar, A. H. Bredekamp, C. Gerber, M. Hegner, and H. P. Lang, Phys. Rev. E. 72 (2005) 031907.
[38] D. W. Chun, K. S. Hwang, K. Eom, J. H. Lee, B. H. Cha, W. Y. Lee, D. S. Yoon, and T. S. Kim, Sens. Actuat. A 135 (2007) 857.
[39] M. Li, H. X. Tang, and M. L. Roukes, Nat. Nanotechnol. 2 (2007) 114.
[40] A. K. Naik, M. S. Hanay, W. K. Hiebert, X. L. Feng, and M. L. Roukes, Nat. Nanotechnol. 4 (2009) 445.
[41] P. A. Greaney and J. C. Grossman, Nano Lett. 8 (2008) 2648.
[42] X. L. Feng, R. He, P. Yang, and M. L. Roukes, Nano Lett. 7 (2007) 1953.
[43] I. Kozinsky, H. W. C. Postma, I. Bargatin, and M. L. Roukes, Appl. Phys. Lett. 88 (2006) 253101.
[44] R. Lifshitz and M. C. Cross, Phys. Rev. B. 67 (2003) 134302.
[45] H. S. Park, W. Cai, H. D. Espinosa, and H. Huang, MRS Bulletin 34 (2009) 178.
[46] L. B. Freund and S. Suresh, Thin Film Materials, Cambridge University Press, Cambridge, 2003.
[47] L. Rayleigh, Phil. Mag. 30 (1890) 285.
[48] R. Shuttleworth, Proc. Phys. Soc. Lond. Sect. A. 63 (1950) 444.
[49] R. E. Miller and V. B. Shenoy, Nanotechnology 11 (2000) 139.
[50] S. Cuenot, C. Fretigny, S. Demoustier-Champagne, and B. Nysten, Phys. Rev. B. 69 (2004) 165410.
[51] J. He and C. M. Lilley, Nano Lett. 8 (2008) 1798.
[52] G. Wang and X. Li, Appl. Phys. Lett. 91 (2007) 231912.
[53] G. Yun and H. S. Park, Phys. Rev. B. 79 (2009) 195421.
[54] H. S. Park, P. A. Klein, and G. J. Wagner, Int. J. Numer. Meth. Engrg. 68 (2006) 1072.
[55] H. S. Park and P. A. Klein, Comput. Meth. Appl. Mech. Engrg. 197 (2008) 3249.
[56] H. S. Park and P. A. Klein, Phys. Rev. B. 75 (2007) 085408.
[57] J. He and C. M. Lilley, Appl. Phys. Lett. 93 (2008) 263108.
[58] A. W. McFarland, M. A. Poggi, M. J. Doyle, L. A. Bottomley, and J. S. Colton, Appl. Phys. Lett. 87 (2005) 053505.
[59] J. Lagowski, H. C. Gatos, and J. E. S. Sproles, Appl. Phys. Lett. 26 (1975) 493.
[60] M. E. Gurtin, X. Markenscoff, and R. N. Thurston, Appl. Phys. Lett. 29 (1976) 529.
[61] H. S. Park, J. Appl. Phys. 103 (2008) 123504.
[62] H. S. Park, Nanotechnology 20 (2009) 115701.
[63] M. J. Lachut and J. E. Sader, Appl. Phys. Lett. 95 (2009) 193505.
[64] K. B. Gavan, H. J. R. Westra, E. W. J. M. van der Drift, W. J. Venstra, and H. S. J. van der Zant, Appl. Phys. Lett. 94 (2009) 233108.

[65] M. C. Cross, J. L. Rogers, R. Lifshitz, and A. Zumdieck, Phys. Rev. E. 73 (2006) 036205.
[66] E. Buks and B. Yurke, Phys. Rev. E. 74 (2006) 046619.
[67] M. D. Dai, K. Eom, and C.-W. Kim, Appl. Phys. Lett. 95 (2009) 203104.
[68] R. He, X. L. Feng, M. L. Roukes, and P. Yang, Nano Lett. 8 (2008) 1756.
[69] S. P. Timoshenko and J. N. Goodier, Theory of elasticity, McGraw-Hill, 1970.
[70] L. Meirovitch, Analytical methods in vibrations, Macmillan, New York, 1967.
[71] J. K. Hale, Oscillations in Nonlinear Systems, McGraw-Hill, New York, 1963.
[72] C. Hayashi, Nonlinear Oscillations in Physical Systems, McGraw-Hill, New York, 1964.
[73] A. H. Nayfeh and D. T. Mook, Nonlinear Oscillations, Wiley, New York, 1979.
[74] C. Q. Chen, Y. Shi, Y. S. Zhang, J. Zhu, and Y. J. Yan, Phys. Rev. Lett. 96 (2006) 075505.
[75] P. Zijlstra, A. L. Tchebotareva, J. W. M. Chon, M. Gu, and M. Orrit, Nano Lett. 8 (2008) 3493.
[76] Y. Zhu, F. Xu, Q. Qin, W. Y. Fung, and W. Lu, Nano Lett. 9 (2009) 3934.
[77] M. J. Gordon, T. Baron, F. Dhalluin, P. Gentile, and P. Ferret, Nano Lett. 9 (2009) 525.
[78] Y.-S. Sohn, J. Park, G. Yoon, S.-W. Jee, J.-H. Lee, S. Na, T. Kwon, and K. Eom, Nanoscale Res. Lett. 5 (2010) 211.
[79] B. Lee and R. E. Rudd, Phys. Rev. B. 75 (2007) 195328.
[80] J. Diao, K. Gall, and M. L. Dunn, Nat. Mater. 2 (2003) 656.
[81] S. S. Verbridge, D. F. Shapiro, H. G. Craighead, and J. M. Parpia, Nano Lett. 7 (2007) 1728.
[82] S. Y. Kim and H. S. Park, Phys. Rev. Lett. 101 (2008) 215502.
[83] S. Y. Kim and H. S. Park, Nanotechnology 21 (2010) 105710.
[84] J. S. Bunch, A. M. van der Zande, S. S. Verbridge, I. W. Frank, D. M. Tanenbaum, J. M. Parpia, H. G. Craighead, and P. L. McEuen, Science 315 (2007) 490.
[85] D. R. Southworth, R. A. Barton, S. S. Verbridge, B. Ilic, A. D. Fefferman, H. G. Craighead, and J. M. Parpia, Phys. Rev. Lett. 102 (2009) 225503.
[86] A. Bokian, Journal of Sound and Vibration 142 (1990) 481.

# Figures Captions

**Figure 1.** Effective elastic modulus of silicon nanowires with respect to their diameters: Here, the elastic modulus of silicon nanowires were extracted from their harmonic resonances that were calculated from our model. The effective elastic modulus was normalized by the elastic modulus of bulk silicon. Moreover, our predictions on the size dependence of elastic modulus for silicon nanowire are compared with experimental data (Zhu et al., *Nano Lett*, 2009) [76] and/or first-principle calculations (Lee et al., *Phys. Rev. B*, 2007) [79].

**Figure 2.** Resonant frequencies for silicon nanowire resonators undergoing the nonlinear vibrations: Inset shows the resonance curves, each of which corresponds to harmonic (blue curve) and nonlinear (red curve) oscillations, respectively. Here, the harmonic resonance was numerically obtained from our model when an actuation force is set to 0.1 fN, while the frequencies for nonlinear oscillation were computed from our model when an actuation force is set to 2.5 pN. Solid lines show the resonant frequencies estimated from our model that takes the surface effect into account, whereas dashed lines indicate the resonant frequencies evaluated from a continuum model that excludes the surface effect.

**Figure 3.** Resonant frequency shifts due to mass adsorption using harmonic oscillations: Inset shows the resonance curves for a bare resonator (blue curve) and a resonator with mass adsorption of 80 ag (red curve), respectively. It is shown that frequency shifts due to mass adsorption due to adsorbed mass depend on the amount of added mass as well as the surface-to-volume ratio of a resonator.

**Figure 4.** Difference between two models – (i) our model that considers the surface effect, and (ii) a continuum model excluding the surface effect – in measuring the resonant frequencies for resonators with mass adsorption: It is found that when small amount of mass (e.g. $\leq$ 10 ag) is added to a resonator, the surface effect does not play a significant role on the frequencies. However, when a sufficient amount of mass (e.g. >20 ag) is adsorbed, the surface effect plays a substantial role in the frequency behavior of nanoresonators that exhibit the large surface-to-volume ratio.

**Figure 5.** Resonant frequency shifts due to mass adsorption for resonators that experiences the nonlinear vibrations: Inset shows the resonance curves corresponding to resonance for a bare resonator, i.e. without mass adsorption (blue curve), and a mass-adsorbed resonator (red curve), respectively. It is shown that mass adsorption increases the resonant frequencies in nonlinear oscillation regime. Frequency shifts due to adsorbed mass for nonlinear resonator is governed by the resonator's surface-to-volume ratio.

**Figure 6.** Difference between two models – (i) our model that considers the surface effect, and (ii) a continuum model excluding the surface effect – in measuring the resonant frequencies for nonlinear resonators with mass adsorption: It is interestingly shown that regardless of amount of adsorbed mass, the surface effect plays a significant role on the frequency behavior of mass-adsorbed resonators that undergo the nonlinear vibrations.

**Figure 7.** Resonant frequency shifts due to mechanical tension applied to a resonator experiencing the harmonic oscillations: Inset shows two resonance curves that correspond to resonance for a resonator without applied tension (blue curve) and a resonator that exerts a mechanical tension (red curve). It is shown that a mechanical tension increases the resonant frequency. It is shown that frequency shifts due to mechanical tension depends on the resonator's surface-to-volume ratio. Here, solid line indicates the frequency shifts computed from our model that accounts for surface effect, while dashed lines represents the frequency shifts calculated from a continuum model excluding the surface effect.

**Figure 8.** Resonant frequency shifts induced by a mechanial tension applied a resonator that undergoes the nonlinear vibration: Inset shows the resonance curves corresponding to a bare resonator (blue curve) and a resonator that bears the mechanical tension (red curve). It is shown that a mechanical tension decreases the resonant frequency in the nonlinear oscillation regime. It is found that mechanical tension significantly reduces the resonant frequencies, and that the decreases in frequencies due to mechanical tension is significantly governed by the resonator's surface-to-volume ratio.

**Figure 9.** Color map that shows the frequency shifts due to mass adsorption (i.e. 40 ag) for resonators that experience the harmonic oscillation as a function of the mechanical tension and the surface-to-volume ratio.

**Figure 10.** Color map indicates the normalized frequency shifts due to mass adsorption (i.e. 40 ag) for resonators that undergo the nonlinear vibrations with respect to the mechanical tension and the surface-to-volume ratio.

**Figures**

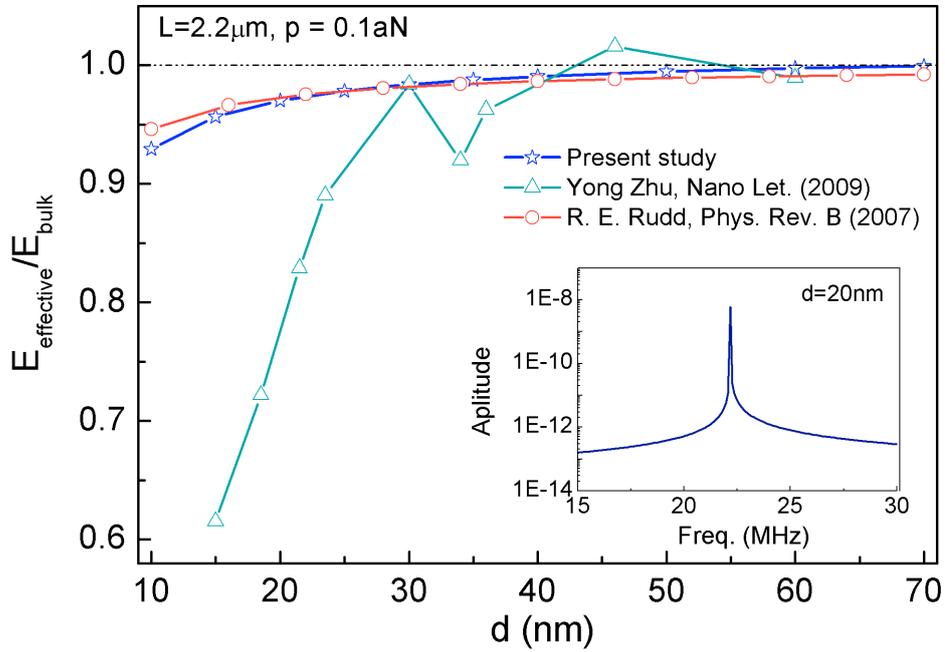

**Figure 1.** Effective elastic modulus of silicon nanowires with respect to their diameters: Here, the elastic modulus of silicon nanowires were extracted from their harmonic resonances that were calculated from our model. The effective elastic modulus was normalized by the elastic modulus of bulk silicon. Moreover, our predictions on the size dependence of elastic modulus for silicon nanowire are compared with experimental data (Zhu et al., *Nano Lett*, 2009) [76] and/or first-principle calculations (Lee et al., *Phys. Rev. B*, 2007) [79].

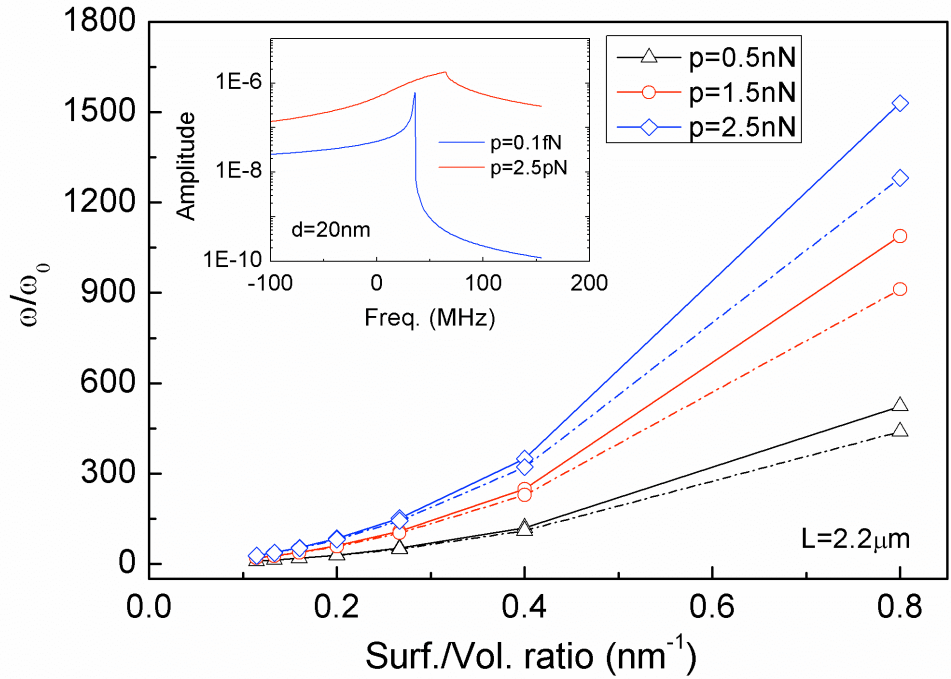

**Figure 2.** Resonant frequencies for silicon nanowire resonators undergoing the nonlinear vibrations: Inset shows the resonance curves, each of which corresponds to harmonic (blue curve) and nonlinear (red curve) oscillations, respectively. Here, the harmonic resonance was numerically obtained from our model when an actuation force is set to 0.1 fN, while the frequencies for nonlinear oscillation were computed from our model when an actuation force is set to 2.5 pN. Solid lines show the resonant frequencies estimated from our model that takes the surface effect into account, whereas dashed lines indicate the resonant frequencies evaluated from a continuum model that excludes the surface effect.

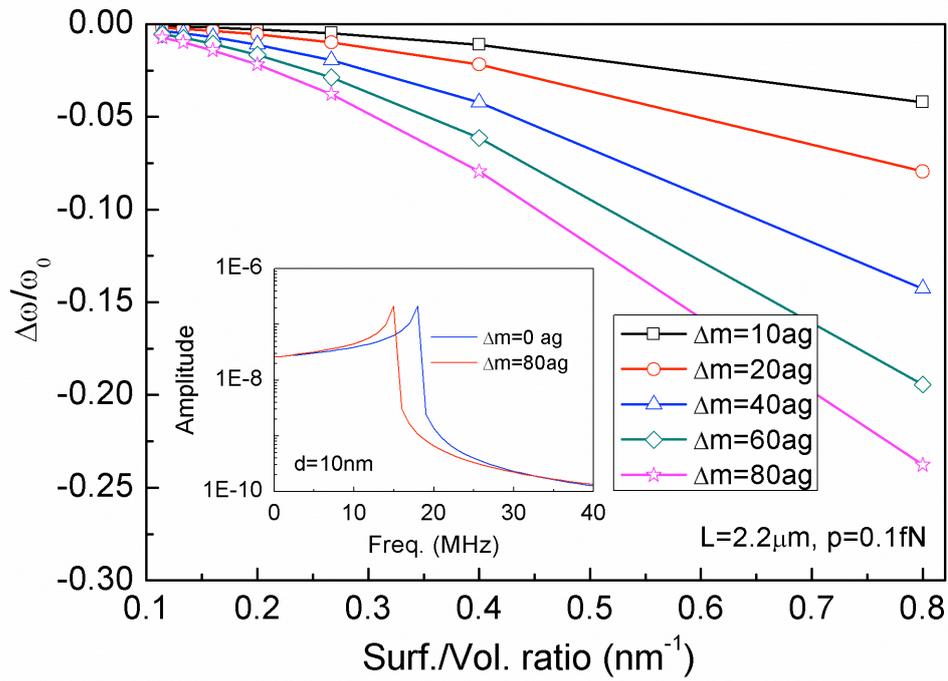

**Figure 3.** Resonant frequency shifts due to mass adsorption using harmonic oscillations: Inset shows the resonance curves for a bare resonator (blue curve) and a resonator with mass adsorption of 80 ag (red curve), respectively. It is shown that frequency shifts due to mass adsorption due to adsorbed mass depend on the amount of added mass as well as the surface-to-volume ratio of a resonator.

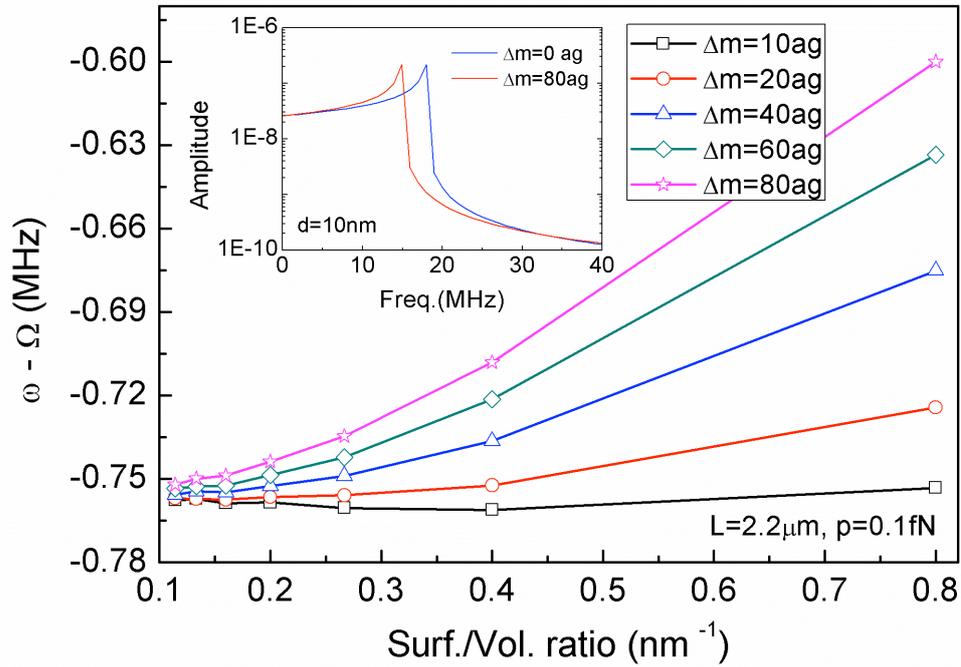

**Figure 4.** Difference between two models – (i) our model that considers the surface effect, and (ii) a continuum model excluding the surface effect – in measuring the resonant frequencies for resonators with mass adsorption: It is found that when small amount of mass (e.g. ≤ 10 ag) is added to a resonator, the surface effect does not play a substantial role on the frequencies. However, when a sufficient amount of mass (e.g. >20 ag) is adsorbed, the surface effect plays a key role in the frequency behavior of nanoresonators that exhibit the large surface-to-volume ratio.

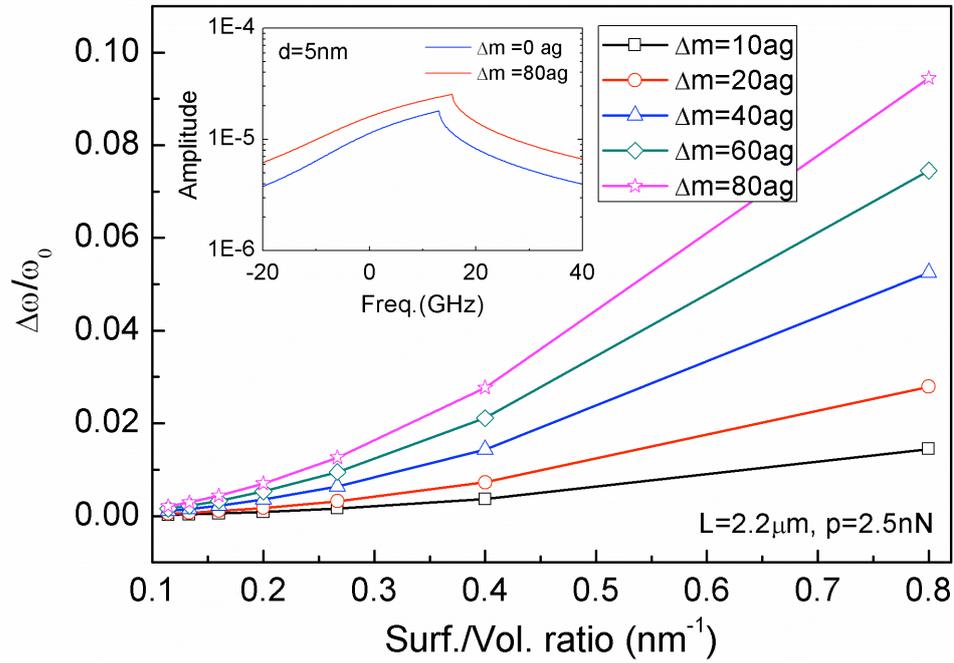

**Figure 5.** Resonant frequency shifts due to mass adsorption for resonators that experiences the nonlinear vibrations: Inset shows the resonance curves corresponding to resonance for a bare resonator, i.e. without mass adsorption (blue curve), and a mass-adsorbed resonator (red curve), respectively. It is shown that mass adsorption increases the resonant frequencies in nonlinear oscillation regime. Frequency shifts due to adsorbed mass for nonlinear resonator is governed by the resonator's surface-to-volume ratio.

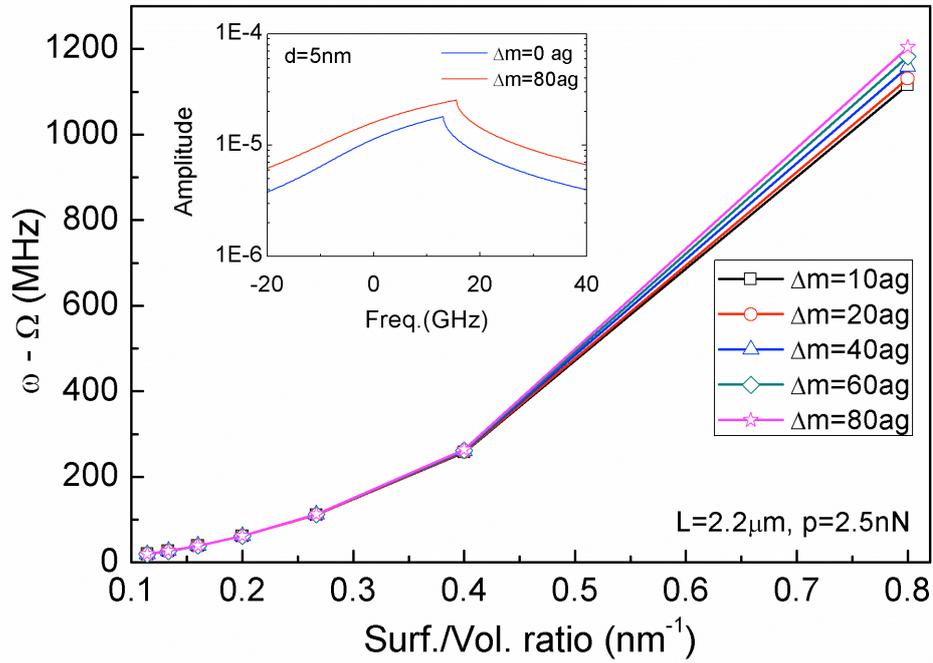

**Figure 6.** Difference between two models – (i) our model that considers the surface effect, and (ii) a continuum model excluding the surface effect – in measuring the resonant frequencies for nonlinear resonators with mass adsorption: It is interestingly shown that regardless of amount of adsorbed mass, the surface effect plays a significant role on the frequency behavior of mass-adsorbed resonators that undergo the nonlinear vibrations.

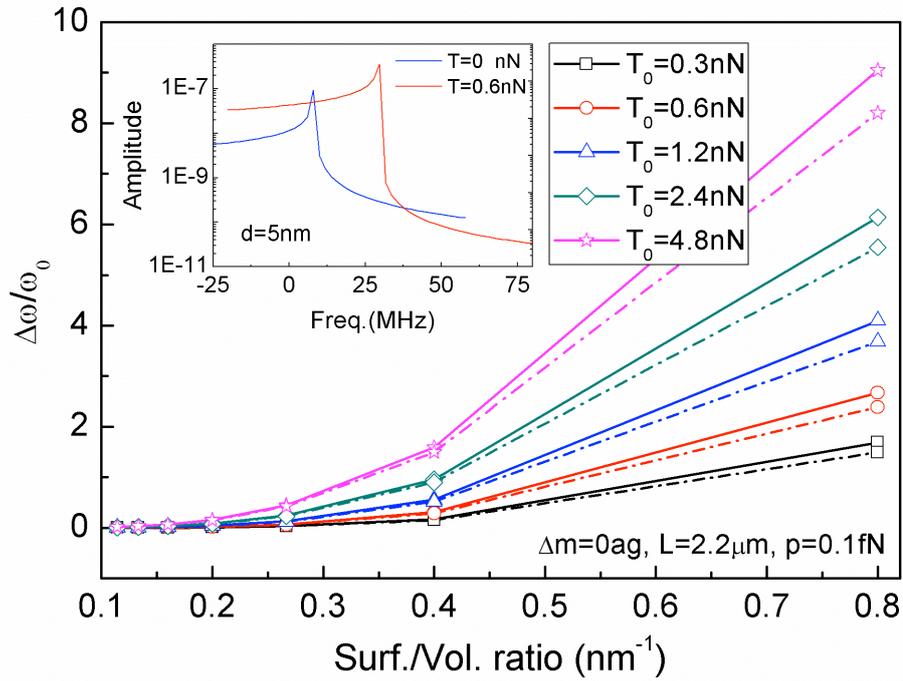

**Figure 7.** Resonant frequency shifts due to mechanical tension applied to a resonator experiencing the harmonic oscillations: Inset shows two resonance curves that correspond to resonance for a resonator without applied tension (blue curve) and a resonator that exerts a mechanical tension (red curve). It is shown that a mechanical tension increases the resonant frequency. It is shown that frequency shifts due to mechanical tension depends on the resonator's surface-to-volume ratio. Here, solid line indicates the frequency shifts computed from our model that accounts for surface effect, while dashed lines represents the frequency shifts calculated from a continuum model excluding the surface effect.

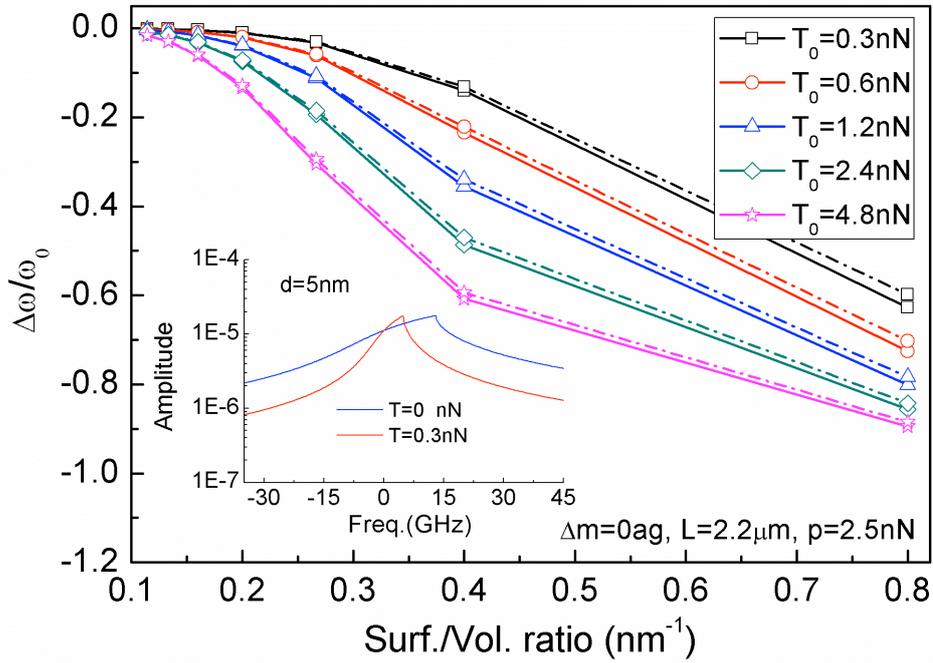

**Figure 8.** Resonant frequency shifts induced by a mechanial tension applied a resonator that undergoes the nonlinear vibration: Inset shows the resonance curves corresponding to a bare resonator (blue curve) and a resonator that bears the mechanical tension (red curve). It is shown that a mechanical tension decreases the resonant frequency in the nonlinear oscillation regime. It is found that mechanical tension significantly reduces the resonant frequencies, and that the decreases in frequencies due to mechanical tension is significantly governed by the resonator's surface-to-volume ratio.

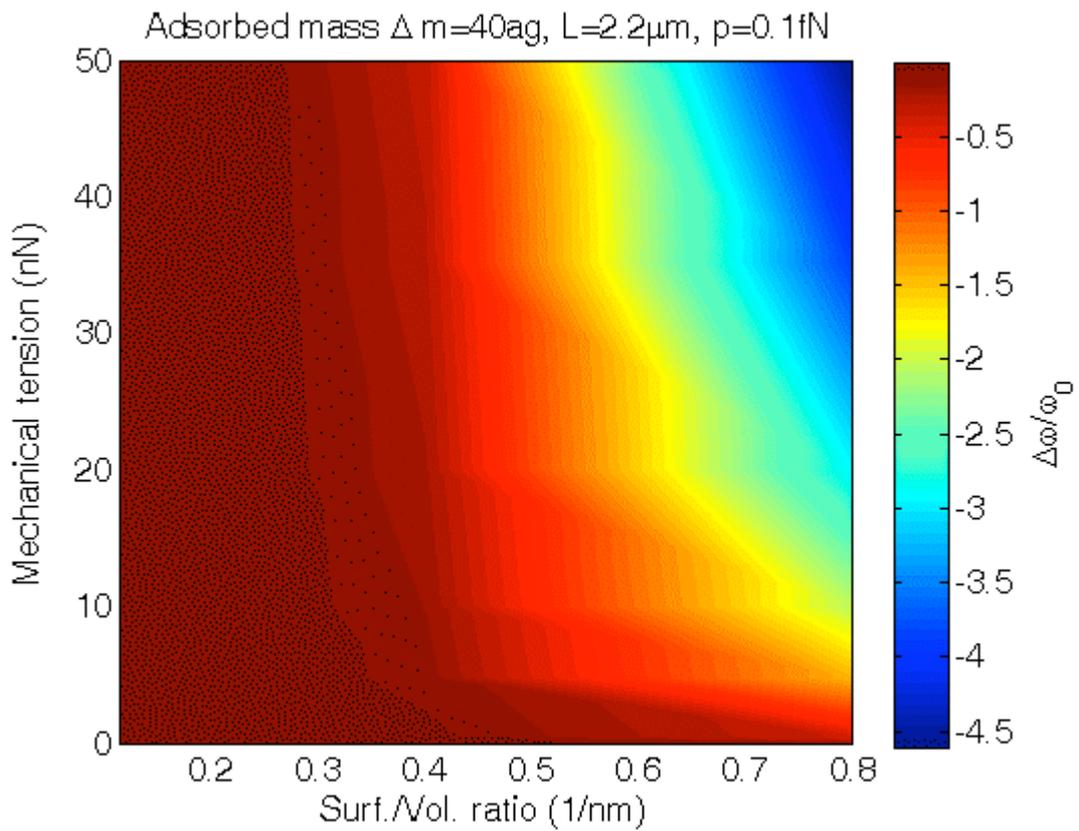

**Figure 9.** Color map that shows the frequency shifts due to mass adsorption (i.e. 40 ag) for resonators that experience the harmonic oscillation as a function of the mechanical tension and the surface-to-volume ratio.

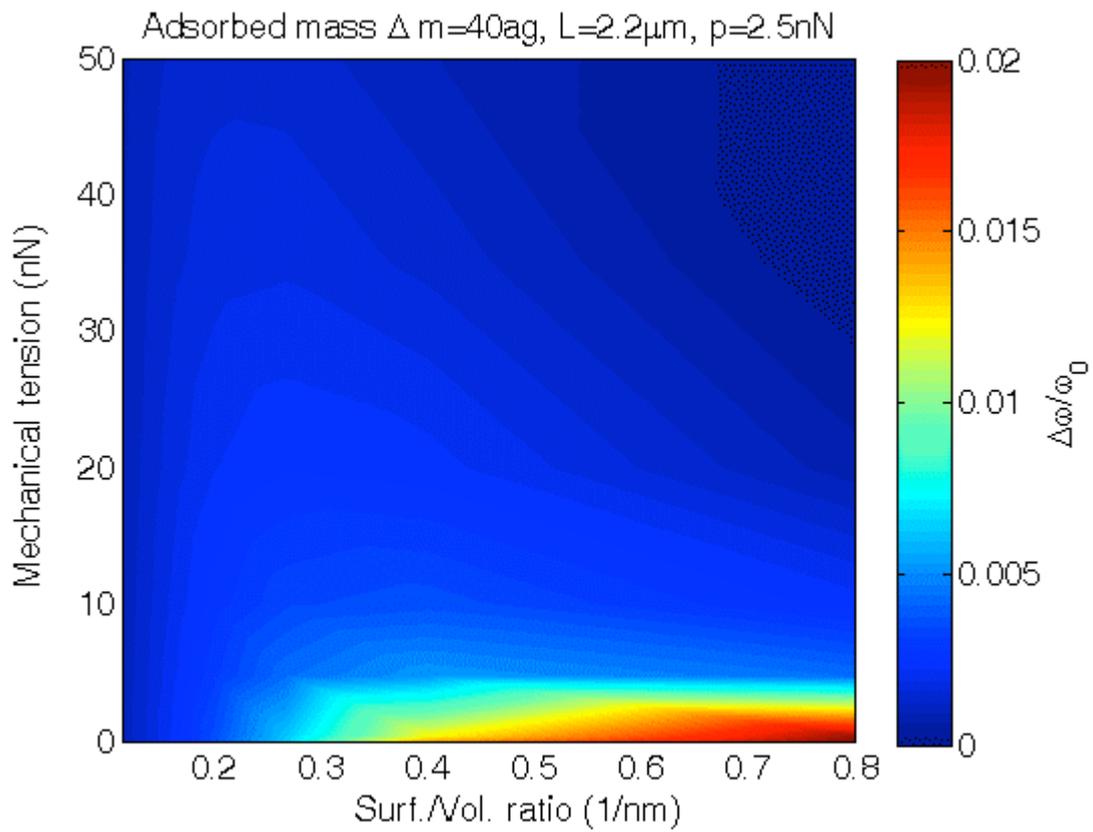

**Figure 10.** Color map indicates the normalized frequency shifts due to mass adsorption (i.e. 40 ag) for resonators that undergo the nonlinear vibrations with respect to the mechanical tension and the surface-to-volume ratio.